\documentstyle[12pt,psfig,epsf]{article}
\newcommand{\be}{\begin{equation}}
\newcommand{\ee}{\end{equation}}

\newcommand{\beqa}{\begin{eqnarray}}
\newcommand{\eeqa}{\end{eqnarray}}

\newcommand{\eqref}[1]{(\ref{#1})}

\def\boxit#1{\vbox{\hrule\hbox{\vrule\kern8pt
\vbox{\hbox{\kern8pt}\hbox{\vbox{#1}}\hbox{\kern8pt}}
\kern8pt\vrule}\hrule}}
\def\mathboxit#1{\vbox{\hrule\hbox{\vrule\kern8pt\vbox{\kern8pt
\hbox{$\displaystyle #1$}\kern8pt}\kern8pt\vrule}\hrule}}

\def\IB{\relax\hbox{$\inbar\kern-.3em{\rm B}$}}
\def\IC{\relax\hbox{$\inbar\kern-.3em{\rm C}$}}
\def\ID{\relax\hbox{$\inbar\kern-.3em{\rm D}$}}
\def\IE{\relax\hbox{$\inbar\kern-.3em{\rm E}$}}
\def\IF{\relax\hbox{$\inbar\kern-.3em{\rm F}$}}
\def\IG{\relax\hbox{$\inbar\kern-.3em{\rm G}$}}
\def\IGa{\relax\hbox{${\rm I}\kern-.18em\Gamma$}}
\def\IH{\relax{\rm I\kern-.18em H}}
\def\IK{\relax{\rm I\kern-.18em K}}
\def\IL{\relax{\rm I\kern-.18em L}}
\def\IP{\relax{\rm I\kern-.18em P}}
\def\IR{\relax{\rm I\kern-.18em R}}
\def\IZ{\relax\ifmmode\mathchoice
{\hbox{\cmss Z\kern-.4em Z}}{\hbox{\cmss Z\kern-.4em Z}}
{\lower.9pt\hbox{\cmsss Z\kern-.4em Z}} {\lower1.2pt\hbox{\cmsss
Z\kern-.4em Z}}\else{\cmss Z\kern-.4em Z}\fi}

\def\II{\relax{\rm I\kern-.18em I}}

\begin{document}

\hfill  NRCPS-HE-02-13

\vspace{5cm}
\begin{center}
{\LARGE  Superstring \\
with\\
Extrinsic Curvature Action\\
}

\vspace{2cm}

{\sl G.K.Savvidy\footnote{email:~savvidy@inp.demokritos.gr}\\
National Research Center Demokritos,\\
Ag. Paraskevi, GR-15310 Athens, Hellenic Republic\\

}
\end{center}
\vspace{60pt}

\centerline{{\bf Abstract}}

\vspace{12pt}

\noindent We suggest supersymmetric extension of
conformally invariant string theory
which is exclusively based on extrinsic curvature action.
At the classical level this is a tension-less string theory.
The absence of conformal anomaly in quantum theory requires
that the space-time should be 6-dimensional.


\newpage

\pagestyle{plain}

A string model which is exclusively based on the concept of extrinsic
curvature was suggested in  \cite{geo}.
It describes random surfaces embedded in D-dimensional
spacetime  with the following action
\be\label{funaction}
S =m \cdot A= {m\over \pi} \int d^{2}\zeta
\sqrt{h}\sqrt{K^{ia}_{a}K^{ib}_{b}},
\ee
where $m$ has dimension of mass, $h_{ab}$ is the induced metric and
$K^{i}_{ab}$ is the second fundamental form (extrinsic curvature).
\footnote{In the above theory extrinsic curvature term
$alone$ should be considered as
fundamental action of the theory. There is no $area$ term in the action
and it is not quadratic in extrinsic curvature
as it was in previous studies \cite{polykov,weingarten}.}
The action (\ref{funaction}) is proportional
to the length of the surface $ A $.
The last property makes the theory very close to
the Feynman path integral for point-like relativistic particle
because when the surface degenerates into a single world line the
action  (\ref{funaction}) becomes  proportional to the length of
the world line
\be\label{limit}
S= m ~A_{xy} ~~ \rightarrow ~~  m~\int^{Y}_{X} ds
\ee
and the functional integral over surfaces
naturally transforms into the Feynman path integral for a point-like
relativistic particle (see Figure 1). For a string which is
stretched between quark-antiquark
pair the action is equal to the perimeter
of the Wilson loop $S=m(R+T)$, where R is space distance between quarks,
therefore at the {\it classical level string tension is equal to zero}.
In the recent articles \cite{geo,manvelyan} the authors demonstrated that
quantum fluctuations generate an area term in
the effective action
\be\label{minimum}
S_{eff} = m^{2}e^{-{4\pi \over D-3}} \int d^{2}\zeta \sqrt{h}
h^{ab}
\partial_{a}X^{\mu}\partial_{b}X^{\mu},
\ee
that is, dynamical string tension
$m^2 exp{(-{4\pi \over D-3})}$.

Our aim now is to introduce fermions and to suggest
supersymmetric extension of this model. First we shall
consider basic relations in the bosonic case and the corresponding
quantization rules of this highly non-linear and higher-derivative theory.
We shall introduce Ramond fermions using standard
two-dimensional world-sheet spinors. The main difference with the
standard superstring theory is that supersymmetry transformation
contains higher derivatives\footnote{The supersymmetric extension
of the model which has the action  quadratic in extrinsic curvature form
was considered in \cite{curtright}.}. We shall demonstrate that
the absence of conformal anomaly in quantum theory requires that
the space-time should be
6-dimensional\footnote{This may be interesting for
little string theory in 6-dimensions\cite{seiberg}.},~$D_c =6$.

\begin{figure}
\centerline{\hbox{\psfig{figure=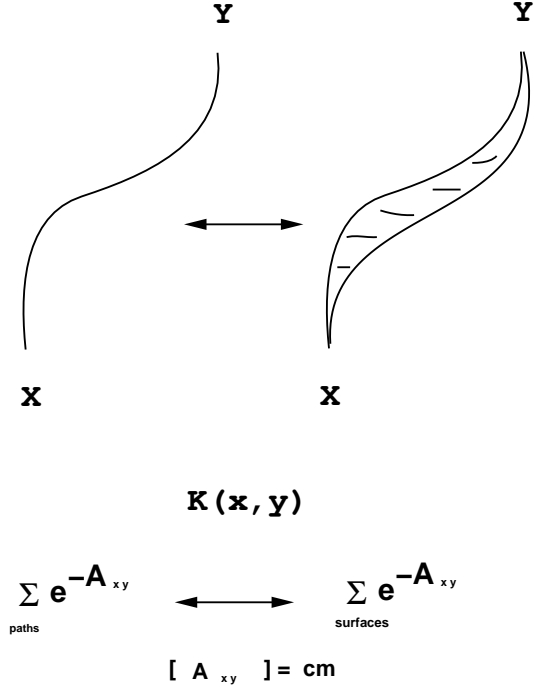,width=7cm}}}
\caption[fig]{ It is required that the action $A_{xy}$ should
measure the surfaces in terms of length, as it was for the path integral.
When a surface degenerates into a single world line we have
natural transition to a path integral .
}
\label{fig}
\end{figure}

We shall represent the gonihedric action
(\ref{funaction}) in the form \cite{geo,manvelyan}
\begin{equation}\label{conaction}
S= {m\over\pi}\int d^{2}\zeta \sqrt{h}\sqrt{ \left(\Delta(h)
X_{\mu}\right)^{2}},
\end{equation}
here ~$h_{ab}=\partial_{a}X_{\mu}\partial_{b}X_{\mu}$ ~is the induced
metric,~$\Delta(h)= 1/\sqrt{h}~\partial_{a}\sqrt{h}h^{ab}
\partial_{b} $ ~is Laplace operator and
$K^{ia}_{a}K^{ib}_{b}=\left(\Delta(h) X_{\mu}\right)^{2}$. The
second fundamental form $K$ is defined through the relations:
$
K^{i}_{ab}n_{\mu}^{i}=\partial_{a}\partial_{b}X_{\mu}-
\Gamma^{c}_{ab}\partial_{c}X_{\mu} \equiv  \nabla_{a}\partial_{b}X_{\mu},
$
where $n_{\mu}^{i}$ are $D-2$ normals and $a,b=1,2; \qquad
\mu=0,1,2,...,D-1; \quad i,j=1,2,...,D-2.$

Below we shall consider a model which has the same
action (\ref{conaction}), but now will be treated  as a functional of
two independent field variables $X^{\mu}$ and $h_{ab}$,
that is, we shall consider scalar fields $X^{\mu}$ in two-dimensional
quantum gravity background $h_{ab}$.
First we shall get equation which follows from the variation
of the action over coordinates $X^{\mu}$
\be\label{massshell}
{\pi \over   \sqrt{h}~}{\delta S \over \delta X^{\mu}} =
\triangle (h)~\left(m {\triangle (h) X^{\mu} \over
\sqrt{(\triangle (h) X)^{2}}} \right) =0.
\ee
and shall introduce the momentum operator
$$P^{\mu}_{a}= \partial_{a}\Pi^{\mu},~~~~~~~~~~~~~~
\Pi^{\mu} =m {\triangle (h) X^{\mu} \over
\sqrt{(\triangle (h) X)^{2}}},
$$
where $\Pi^{\mu}$ has the property very similar with
the constraint equation
for the point-like relativistic particle
\be\label{secondcon}
\Pi^{\mu}\Pi_{\mu}~ =~ m^{2}.
\ee
One should also compute the variation of the action with respect to the metric
$h^{ab}$
\be
\delta S = - {1\over 2\pi} \int \sqrt{h} T_{ab} \delta h^{ab} d^{2}\zeta =0
\ee
with the following result for $T_{ab}$
\be\label{tensor}
T_{ab} = \nabla_{\{a }\Pi^{\mu}  ~\nabla_{b\}} X^{\mu}
~-~h_{ab}~h^{cd} ~ \nabla_{c}~ \Pi^{\mu}~\nabla_{d} X^{\mu}~=~0,
\ee
where $\{a ~~~b\} $ denotes a symmetric sum.  The energy-momentum
tensor is traceless $h^{ab} T_{ab} =0$ and we have interaction
with conformally
invariant matter field $X^{\mu}$. Thus our basic equations are
(\ref{massshell}),(\ref{secondcon}) and (\ref{tensor}).

We can fix the conformal gauge $h_{ab}=\rho\eta_{ab}$ using
reparametrization invariance of the action and to derive it in the form
(see (\ref{conaction}))
\be\label{gaga}
\acute{S} = {m\over\pi} \int d^{2}\zeta \sqrt{\left(\partial^{2}
X \right)^{2}} ~=~{1\over\pi}\int d^{2}\zeta ~\Pi^{\mu}~\partial^{2} X^{\mu}.
\ee
In this gauge the equation of motion (\ref{massshell})
should be accompanied by the constraint equations (\ref{tensor}) where
now
$$\Pi^{\mu} = m \frac{\partial^{2}X^{\mu}}
{\sqrt{\left(\partial^{2}X \right)^2}}.
$$
In the light cone coordinates
$\zeta^{\pm}=\zeta^{0}\pm\zeta^{1}$,
$\partial_{\pm}= {1\over 2}(\partial_{0} \pm \partial_{1})$
the constrains (\ref{tensor}) take the form
\beqa\label{cons9}
T_{++}= {1\over 2} ~(T_{00} + T_{01})=
{1\over 2}(\partial_{0}\Pi^{\mu} +\partial_{1}\Pi^{\mu})
(\partial_{0}X^{\mu} +\partial_{1}X^{\mu})= 2~\partial_{+}\Pi^{\mu}
\partial_{+}X^{\mu},\nonumber\\
T_{--}= {1\over 2} ~(T_{00} - T_{01})= {1\over 2}
(\partial_{0}\Pi^{\mu} - \partial_{1}\Pi^{\mu})
(\partial_{0}X^{\mu} - \partial_{1}X^{\mu})= 2~\partial_{-}\Pi^{\mu}
\partial_{-}X^{\mu},
\eeqa
with the trace equal to zero: $T_{+-}=0$.
The conservation of the energy momentum tensor takes the form
$
\partial_{-}T_{++} = \partial_{+}T_{--}=0
$
and  requires that its components are analytic $T_{++}=T_{++}(\zeta^{+})$
and anti-analytic $~T_{--}=T_{--}(\zeta^{-})$ functions. Thus our
system has infinite number of conserved charges.
This residual symmetry can be easily   seen in gauge
fixed action (\ref{gaga}) written in light cone coordinates
$
\acute{S} = {4m\over\pi} \int \sqrt{(\partial_{+}\partial_{-}
X^{\mu})^{2}}~ d\zeta^{+}d\zeta^{-},
$
it is invariant under the transformations
$
\zeta^{+} = f({\tilde{\zeta}^{+}}),~~~\zeta^{-} =
g({\tilde{\zeta}^{-}})
$
where $f$ and $g$ are arbitrary functions. Another important
symmetry of the equation (\ref{massshell}) is the transformation
$
\partial^{2}X^{\mu} \rightarrow
\Omega(\zeta^{+},\zeta^{-})~ \partial^{2}X^{\mu},
$
where $\Omega(\zeta^{+},\zeta^{-})$ is an arbitrary function.
Indeed if $\partial^{2}X^{\mu}$ is a solution of the equation
(\ref{massshell}), then $\partial^{2}\Phi^{\mu} =
\Omega(\zeta^{+},\zeta^{-})~ \partial^{2}X^{\mu} $ is also
a solution.

Let us also consider additional conserved currents.
If $\delta X^{\mu}$ is symmetry transformation of the
action with only second derivatives of the fields, then
the corresponding conserved current is :
\beqa
J_{a} = \partial_{a}\left({\partial L \over \partial(\partial^{2}X^{\mu}) }\right)
\delta X^{\mu} - {\partial L \over \partial(\partial^{2}X^{\mu}) }
\partial_{a}\delta X^{\mu}. \nonumber
\eeqa
In our case it is equal to:
$
J_{a} = \partial_{a}\Pi_{\mu} \cdot \delta X^{\mu} -
\Pi_{\mu}  \cdot\partial_{a}\delta X^{\mu}.
$
The action (\ref{gaga}) is invariant under the global symmetries
$\delta X^{\mu} =\Lambda^{\mu\nu}X_{\nu} + a^{\mu}$,
where $\Lambda^{\mu\nu}$ is a constant antisymmetric matrix while
$a^{\mu}$ is a constant. The translation invariance of the action (\ref{gaga})
$\delta_{a} X^{\mu} = a^{\mu}$ results into the conserved momentum
current (already appearing in (\ref{massshell}))
\be
P^{\mu}_{a} = \partial_{a}\Pi^{\mu},~~~~~~~\partial^{a}
P^{\mu}_{a} =0,~~~~P^{\mu} = \int P^{\mu}_0 d\zeta^1
\ee
and Lorentz transformation
$\delta_{\Lambda} X^{\mu} = \Lambda^{\mu\nu}X_{\nu}$ into angular momentum
current
\be
M^{\mu\nu}_{a} = X^{\mu}\partial_{a}\Pi^{\nu} - X^{\nu}\partial_{a}\Pi^{\mu}
+ \Pi^{\mu}\partial_{a}X^{\nu} - \Pi^{\nu}\partial_{a}X^{\mu},~~~~~~~\partial^{a}
M^{\mu\nu}_{a} =0,~~~~M^{\mu\nu} = \int M^{\mu\nu}_0 d\zeta^1
\ee
In this set up the conformal transformation
$
\zeta^{+} = f({\tilde{\zeta}^{+}}),~~~\zeta^{-} =
g({\tilde{\zeta}^{-}})
$ is defined as
$\delta_{c} X^{\mu} = u^{a}\partial_{a}X^{\mu}$, where $u^{0} = f+g,~u^{1} = f-g $
and the corresponding conserved current is $T_{ab}$ (\ref{tensor}).
We can now summarize the basic equations, constraints and boundary conditions for
closed string in the form:
\beqa
\partial^{2}\Pi^{\mu} =0,~~~~~~~~~~
\Pi^{\mu}= m \frac{\partial^{2}X^{\mu}}
{\sqrt{\left(\partial^{2}X_{\mu}\right)^2}},~~~~~~\Pi^{2}_{\mu} =m^2 ,\nonumber\\
T_{++}= \partial_{+}\Pi^{\mu}
\partial_{+}X^{\mu}=0,\nonumber~~~~~~~~~~~~~~~~~~~~~~~~~~~~~~~~~~~~~~~\\
T_{--}= \partial_{-}\Pi^{\mu}
\partial_{-}X^{\mu}=0,~~~~~~~~~~~~~~~~~~~~~~~~~~~~~~~~~~~~~~~\nonumber\\
X^{\mu}(\zeta^{0},\zeta^{1}) = X^{\mu}(\zeta^{0},\zeta^{1}+2\pi).
~~~~~~~~~~~~~~~~~~~~~~~~~~~~~~~~~
\eeqa
The general solution of the equation (\ref{massshell}) can be
represented in the form \cite{lastprep}
$$
\Pi^{\mu} = {1 \over 2} ~\Pi^{\mu}_{R} +
{1 \over 2} ~\Pi^{\mu}_{L},~~~~X^{\mu} = {1 \over 2}~ X^{\mu}_{R} +
{1 \over 2} ~X^{\mu}_{L} + {1\over 2}
\int[\Pi^{\mu}_{L} +
\Pi^{\mu}_{R}]~\Omega d\zeta^+d\zeta^-,
$$
where $\Omega(\zeta^+ ,\zeta^-)$ is arbitrary function
of $\zeta^+$ and $\zeta^-$.
From its definition the momentum density,~~
$
P^{\mu}(\zeta^{0},\zeta^{1})
\equiv P^{\mu}_{0}(\zeta^{0},\zeta^{1}) =
\partial_{0}\Pi^{\mu}
$
is conjugate to ~$X^{\mu}(\zeta^{0},\zeta^{1})$,~~ therefore
$
[X^{\mu}(\zeta^{0},\zeta^{1}), P^{\nu}(\zeta^{0},\zeta^{'1})]
= i\eta^{\mu\nu} \delta(\zeta^{1} - \zeta^{'1})
$
and one can deduce that the following commutation relations should hold
\beqa
[\partial_{+} X^{\mu}_{L}(\zeta),
\partial_{+}\Pi^{\nu}_{L}(\zeta^{'})]= i \pi \eta^{\mu\nu} \delta^{~'}
(\zeta  - \zeta^{'}),\nonumber\\
~[\partial_{-} X^{\mu}_{R}(\zeta),
\partial_{-}\Pi^{\nu}_{R}(\zeta^{'})]= i \pi \eta^{\mu\nu} \delta^{~'}
(\zeta  - \zeta^{'}),
\eeqa
with all others equal to zero ~$[\partial_{\pm} X^{\mu}_{L}(\zeta),
\partial_{\pm}X^{\nu}_{L}(\zeta^{'})]=0$,~~$[\partial_{\pm} \Pi^{\mu}_{L}(\zeta),
\partial_{\pm}\Pi^{\nu}_{L}(\zeta^{'})]=0$\footnote{Requiring that commutation
relations for the energy-momentum tensor $T_{++}$ and $T_{--}$ should
form the algebra of the two-dimensional conformal group one can
get the same form of basic commutators.}. To make these formulas
more transparent to the reader let me use
the analogy with the ghosts $c^{\pm}$ and anti-ghost $b_{\pm\pm}$ fields
(or super-ghorts). The standard Faddeev-Popov action has the
form $\int c~\partial_{\pm} b$ with
nonzero anti-commutator only between $c$ and $b$ fields.
Making use of these commutators one can get the
algebra of constraints
\beqa\label{standardalg}
[T_{++}(\zeta), T_{++}(\zeta^{'})]=
i \pi(T_{++}(\zeta) + T_{++}(\zeta^{'}))~\delta^{~'}
(\zeta - \zeta^{'}),~~~~~~~\nonumber\\
~[T_{++}(\zeta), P^{2}_{+}(\zeta^{'})]=
i \pi P^{2}_{+}(\zeta)~\delta^{~'}
(\zeta - \zeta^{'}),~~~[P^{2}_{+}(\zeta), P^{2}_{+}(\zeta^{'})]=0,
\eeqa
with similar relations for $T_{--}$ and $P_{-}$.
Here $2P^{\mu}_{0} = \partial_{+}\Pi^{\mu}_{L} + \partial_{-}\Pi^{\mu}_{R}
= P^{\mu}_{+} + P^{\mu}_{-}$.

Our aim now is to include fermions.
We shall introduce fermions into string theory with
extrinsic curvature action using standard
two-dimensional world-sheet Majorana spinors
\cite{ramond,neveu,zumino,green,polchinski}
\be
\Psi^{\mu}_{A}(\zeta) \equiv \left( \begin{array}{c}
     \Psi^{\mu}_{-}(\zeta)\\
     \Psi^{\mu}_{+}(\zeta)
     \end{array} \right),
\ee
where $\mu$ is a space time vector index, $A=1,2$ is a
two-dimensional spinor index.
The action is a sum of the bosonic  and fermionic terms
\be
\acute{S} = {m\over\pi} \int d^{2}\zeta \{ \sqrt{\left(\partial^{2}
X^{\mu}\right)^{2}} + i~\bar{\Psi}^{\mu} \rho^{a}\partial_{a}\Psi^{\mu} \}
\ee
where $\bar{\Psi}^{\mu} = \Psi^{+\mu} \rho^{0}= \Psi^{T\mu} \rho^{0}$ and
$\rho^{\alpha}$ are two-dimensional Dirac matrices
\be
\{ \rho^{a},\rho^{b} \} =-2 \eta^{a b}.
\ee
In Majorana basis the $\rho's$ are give by
\be
\rho^{0} = \left( \begin{array}{cc}
     0&-i\\
     i&0
     \end{array} \right),~~~~\rho^{1} = \left( \begin{array}{cc}
     0&i\\
     i&0
     \end{array} \right)
\ee
and $i\rho^{\alpha}\partial_{\alpha}$ is a real operator.
The two-dimensional chiral fields are defined as
$
\rho^{3}\Psi^{\mu}_{\pm} = \mp \Psi^{\mu}_{\pm},
$
where $\rho^{3}=\rho^{0}\rho^{1}$,
and our field equations can be written in a compact form
\be
\partial_{+}\partial_{-}\Pi^{\mu}_{R}=0 \Leftrightarrow
\partial_{+}\Psi^{\mu}_{-}=0,
\ee
\be
\partial_{-}\partial_{+}\Pi^{\mu}_{L}=0 \Leftrightarrow
\partial_{-}\Psi^{\mu}_{+}=0.
\ee
The symmetry transformation is:
\beqa\begin{array}{ll}\label{trans}
\delta X^{\mu} = \bar{\epsilon}\Psi^{\mu},\\
\delta \Psi^{\mu} = -i \rho^{a} \partial_{a} \Pi^{\mu} ~\epsilon,
\end{array}
\eeqa
where the anti-commuting parameter
$\epsilon$ is a two-dimensional spinor
\be
\epsilon \equiv \left( \begin{array}{c}
     \epsilon_{-}\\
     \epsilon_{+}
     \end{array} \right).
\ee
and $P^{\mu}_{a} =\partial_{a}\Pi^{\mu}$ is the momentum operator.
This transformation does leave the set of field equations
\be
\partial^{2}~\Pi^{\mu}=0
,~~~~~~~\rho^{a}\partial_{a}\Psi^{\mu}=0
\ee
intact. Indeed:
\be
\rho^{a}\partial_{a}\delta\Psi^{\mu}=\rho^{a}\partial_{a}
(-i \rho^{b} \partial_{b} \Pi^{\mu} \cdot \epsilon )=
-i\rho^{a}\rho^{b}\partial_{a}\partial_{b}\Pi^{\mu} \cdot\epsilon=
i\partial^{2} \Pi^{\mu} \cdot\epsilon =0
\ee
\be
\partial^{2} \delta\Pi^{\mu} = \partial^{2}
\{ ~{\eta^{\mu\nu}- \Pi^{\mu}\Pi^{\nu}   \over \sqrt{(\partial^{2} X)^{2}}  }
~\bar{\epsilon} ~\partial^{2} \Psi^{\nu} ~\}=0
\ee
and field equations are invariant under transformation (\ref{trans}).

We should now test whether we have closed supersymmetry  algebra
of transformations (\ref{trans}).
The commutator of two supersymmetries on $X^{\mu}$ is given by
\be\label{susy}
[\delta_{1},\delta_{2}]X^{\mu} = \delta_{1}
\bar{\epsilon}_{2}\Psi^{\mu} - \delta_{2}
\bar{\epsilon}_{1}\Psi^{\mu} = \bar{\epsilon}_{2}
(-i\rho^{a}\partial_{a}\Pi^{\mu}~\epsilon_{1}) -
\bar{\epsilon}_{1}(-i\rho^{a}\partial_{a}\Pi^{\mu}~\epsilon_{2})
= 2i\bar{\epsilon}_{1} \rho^{a}\epsilon_{2} ~ \partial_{a}\Pi^{\mu},
\ee
where the relation $\bar{\epsilon}_{2} \rho^{a}\epsilon_{1}=
-\bar{\epsilon}_{1} \rho^{a}\epsilon_{2}$ has been used.
The commutator of two supersymmetries on $\Psi^{\mu}$ is given by
\be\label{susy1}\begin{array}{lrl}
\begin{array}{l}
[\delta_{1},\delta_{2}]\Psi^{\mu} = \delta_{1}
(-i\rho^{a}\partial_{a}\Pi^{\mu}~\epsilon_{2}) -
\delta_{2}(-i\rho^{a}\partial_{a}\Pi^{\mu}~\epsilon_{1})=
\end{array}\\\\
\begin{array}{r}
-i\rho^{a}\partial_{a} \{{\eta^{\mu\nu}- \Pi^{\mu}\Pi^{\nu}
 \over \sqrt{(\partial^{2} X)^{2}}  }~
\bar{\epsilon}_{1} ~\partial^{2} \Psi^{\nu} \}~\epsilon_{2} \Leftrightarrow
(1\leftrightarrow 2)=
\end{array}\\\\
\begin{array}{l}
= 2i\bar{\epsilon}_{1} \rho^{b}\epsilon_{2} ~ \partial_{b}
\{{\eta^{\mu\nu}- \Pi^{\mu}\Pi^{\nu}
 \over \sqrt{(\partial^{2} X)^{2}}  }
~\partial^{2} \Psi^{\nu} \} +
i\bar{\epsilon}_{1} \rho^{b}\epsilon_{2} ~\rho_{b}~ \rho^{a}
\partial_{a}\{{\eta^{\mu\nu}- \Pi^{\mu}\Pi^{\nu}
 \over \sqrt{(\partial^{2} X)^{2}}  }
~\partial^{2} \Psi^{\nu} \} =0.
\end{array}
\end{array}
\ee
Last two terms are equal to zero on mass shell because of the
classical equation $\partial^{2}\Psi^{\mu}=0$. The above
calculation makes use of the Fierz rearrangement as well as the
properties of Majorana spinors.

It is easy to prove that the action is invariant under the
transformation of equation (\ref{trans}). The invariance
is achieved without the use of the field equations,
\beqa
\delta S = \int \{ \Pi^{\mu} \partial^{2}\delta X^{\mu}
+ 2i \bar{\Psi}^{\mu}\rho^{a}\partial_{a}\delta\Psi^{\mu}  \}
d^{2}\zeta = \int \{ \Pi^{\mu} \partial^{2} (\bar{\epsilon} \Psi^{\mu})
+ 2i \bar{\Psi}^{\mu}\rho^{a}\partial_{a}(-i\rho^{b}\partial_{b}
\Pi^{\mu} \epsilon) \} d^{2}\zeta \nonumber\\=
\int \{ \partial^{2} \Pi^{\mu} (\bar{\epsilon} \Psi^{\mu})
+ 2 \bar{\Psi}^{\mu}\rho^{a}\rho^{b} \partial_{a}\partial_{b}
\Pi^{\mu} \epsilon) \} d^{2}\zeta =
\int \{ \partial^{2} \Pi^{\mu} ~\bar{\epsilon} \Psi^{\mu}
- \bar{\Psi}^{\mu}\epsilon ~\partial^{2}
\Pi^{\mu} \} d^{2}\zeta =0
\eeqa
We can use the Noether method to derive conserved supersymmetry
current, the current is a spinor
\be
J_{a} = {1\over 2 }\rho^{b}\rho_{a}
\Psi^{\mu}\partial_{\beta}\Pi^{\mu} \equiv (J_+ ,J_-),
\ee
here $J_+$ and $J_-$ are two component spinor currents. The current is
conserved
\be
\partial^{a} J_{a} ={1\over 2 }\rho^{b}\rho_{a}
\partial^{a}\Psi^{\mu}~\partial_{b}\Pi^{\mu} +
{1\over 2 }\rho^{b}\rho_{a}
\Psi^{\mu}\partial^{a}\partial_{b}\Pi^{\mu} =0
\ee
and
\be
\rho^{a}J_{a} ={1\over 2 }\rho^{a}\rho^{b}\rho_{a}
\Psi^{\mu}\partial_{b}\Pi^{\mu}=0
\ee
because $\rho^{a}\rho^{b}\rho_{a}=0$.
In the two-dimensional notations the supercurrent has components:
\beqa
J_+ = \Psi^{\mu}_{+}\partial_{+}\Pi^{\mu},~~~~~\partial_{-} J_+ =0,\\
J_- = \Psi^{\mu}_{-}\partial_{-}\Pi^{\mu},~~~~~\partial_{+} J_+ =0,
\eeqa
representing the {\it generalized Dirac equations} in our case.

Let us summarize the symmetries of the system. They are:
translation in the target space $\delta_{a}$,
spacetime rotations $\delta_{\Lambda}$, fermi-bose
transformation $\delta_{\epsilon}$ and conformal transformations
$\delta_{u}$
\be\label{symmetries}\begin{array}{lll}
\delta_{a}:~~~~~~\begin{array}{ll}
\delta_{a}X^{\mu}=a^{\mu}& ~~~~~~~~~~~~~~~P^{\mu}_{a}=\partial_{a}\Pi^{\mu}\\
\delta_{a}\Psi^{\mu} =0
\end{array}\\ \\
\delta_{\Lambda}:~~~~~~\begin{array}{ll}
\delta_{\Lambda}X^{\mu}=\Lambda^{\mu\nu}X^{\nu}& ~~~~~~~M^{\mu\nu}_{a}
=X^{\mu}\partial_{a}\Pi^{\nu}-X^{\nu}\partial_{a}\Pi^{\mu}+
\Pi^{\mu}\partial_{a}X^{\nu} - \Pi^{\nu}\partial_{a}X^{\mu}
+ i\Psi^{\mu}\rho_{a}\Psi^{\nu} \\
\delta_{\Lambda}\Psi^{\mu} =\Lambda^{\mu\nu}\Psi^{\nu}
\end{array}\\
\\
\delta_{\epsilon}:~~~~~~\begin{array}{ll}
\delta_{\epsilon}X^{\mu}=\bar{\epsilon}\Psi^{\mu}& ~~J_{a} =
{1\over 2 }\rho^{b}\rho_{a}\Psi^{\mu}\partial_{b}\Pi^{\mu}  \\
\delta_{\epsilon}\Psi^{\mu} =
-i \rho^{a} \partial_{a} \Pi^{\mu}\cdot\epsilon
\end{array}\\\\
\delta_{u}:~~~~~~\begin{array}{ll}
\delta_{u}X^{\mu}=u^{a}\partial_{a}X^{\mu}&~~~~~~~T_{a b} =
{1\over 2 }\partial_{\{a}\Pi^{\mu}\partial_{b\}}X^{\mu}  +
{i\over 4 }\bar{\Psi}^{\mu}\rho_{\{a}\partial_{b\}}\Psi^{\mu}-trace\\
\delta_{u}\Psi^{\mu} = u^{a}\partial_{a}\Psi^{\mu}
\end{array}\\
\end{array}
\ee
Now we can clearly see that two world sheet supersymmetry transformations
are equivalent to translation $\delta_a$ in (\ref{symmetries})
\beqa
\int_{0}^{2\pi} [\delta_{1},\delta_{2}]X^{\mu} d\zeta^1 =
2i\bar{\epsilon}_{1} \rho^{a}\epsilon_{2} ~ \int_{0}^{2\pi}
\partial_{a}\Pi^{\mu} d\zeta^1 = 2i\bar{\epsilon}_{1} \rho^{0}\epsilon_{2}
~ \int_{0}^{2\pi} \partial_{0}\Pi^{\mu} d\zeta^1 =
2i\bar{\epsilon}_{1} \rho^{0}\epsilon_{2} ~P^{\mu},\\
\int_{0}^{2\pi} [\delta_{1},\delta_{2}]\Psi^{\mu} d\zeta^1 = 0.
~~~~~~~~~~~~~~~~~~~~~~~~~~~~~~~~~~~~
\eeqa
Normally in standard string theory it
is equivalent to a worldsheet conformal transformation.

Defining the conjugate variable for the fermion field as
$
P^{\mu}_{\psi} = {\delta S\over \delta\partial_{0}\Psi^{\mu}} = \Psi^{\mu}
$
we can get standard anticommutation relations
\beqa\begin{array}{lll}
\{ \Psi^{\mu}_{\pm}(\zeta),\Psi^{\nu}_{\pm}(\zeta^{'})\}= \pi
\eta^{\mu\nu} \delta(\zeta^{1} - \zeta^{'}),\\
\end{array}
\eeqa
with all others equal to zero.
Making use of these commutators one can get the standard
algebra of the two-dimensional conformal group (\ref{standardalg}),
where now
\beqa
T_{++}=2\partial_{+}\Pi^{\mu} \partial_{+}X^{\mu} +
{i\over 2 } \Psi^{\mu}_{+} \partial_{+}\Psi^{\mu}_{+},   \nonumber\\
T_{--}=2\partial_{-}\Pi^{\mu} \partial_{-}X^{\mu}+
{i\over 2 } \Psi^{\mu}_{-} \partial_{-}\Psi^{\mu}_{-}.
\eeqa
We are able now to
compute the anticommutator  of two supercurrents:
\beqa
\{ J_{+}(\zeta),J_{+}(\zeta^{'})\} =
\{ \Psi^{\mu}_{+}(\zeta),\Psi^{\nu}_{+}(\zeta^{'})\}\cdot
\partial_{+}\Pi^{\mu}(\zeta)\partial_{+}\Pi^{\nu}(\zeta^{'})
\equiv
\pi P^{2}_{+}(\zeta)~\delta(\zeta^{1} - \zeta^{'}).
\eeqa
The square of our generalized Dirac operator
$J_{+}=P^{\mu}_{+}~\Psi^{\mu}_{+} = \partial_{+}\Pi^{\mu} ~\Psi^{\mu}_{+}$ is not any more
equal to the Virasoro operator $T_{++}= 2 P^{\mu}_{+}\partial_{+}X^{\mu}~+~
(i/2)\Psi^{\mu}_{+}\partial_{+}\Psi^{\mu}_{+}$,
but instead is a square of  space-time translation operator $P^{\mu}_{+}$.
One can also compute the following commutator
\be
[T_{++}(\zeta), J_{+}(\zeta^{'})]= i \pi J_{+}(\zeta)~\delta^{~'}
(\zeta - \zeta^{'}).
\ee

Quantization of this theory is strait forward \cite{lastprep}.
From the equivalent form of the action (\ref{gaga})
we can deduce the propagator
\be\label{propagator}
\langle \Pi^{\mu}(k)X^{\nu}(-k)\rangle
=\eta^{\mu\nu}{i\pi\over k^2},
\ee
or in the coordinate form
\be
<\Pi^{\mu}(\zeta)X^{\nu}(\tilde{\zeta})> = -{\eta^{\mu\nu}\over 2}
ln (\vert \zeta - \tilde{\zeta} \vert \mu).
\ee
Using the fact that there is no correlations between
right and left moving modes of the $\Pi$ field
and right and left moving modes of the $X$ field we shall get
\be\label{contraction}
<\Pi^{\mu}_{R}(\zeta^-)X^{\nu}_{R}(\tilde{\zeta^-})> =
~-\eta^{\mu\nu}~ln [( \zeta^{-} - \tilde{\zeta}^{-})\mu)],\\
<\Pi^{\mu}_{L}(\zeta^+)X^{\nu}_{L}(\tilde{\zeta^+})> =
~ -\eta^{\mu\nu}~ln [( \zeta^{+} - \tilde{\zeta}^{+})\mu)]
\ee
Now we are in a position to compute the two point correlation function
of the energy momentum operator for bosonic coordinates $X,\Pi$
\beqa
<T~T^{boson}_{++}(\zeta^{+}) ~T^{boson}_{++}(\tilde{\zeta}^{+})  >~ =~
{1\over 4} <T:\dot{\Pi}^{\mu}_{L}(\zeta^{+}) \dot{X}^{\mu}_{L}(\zeta^{+}):
:\dot{\Pi}^{\nu}_{L}(\tilde{\zeta}^{+})
\dot{X}^{\nu}_{L}(\tilde{\zeta}^{+}):>~ \nonumber\\
=~{1\over 4}   <\dot{\Pi}^{\mu}_{L}(\zeta^{+})
\dot{X}^{\nu}_{L}(\tilde{\zeta}^{+}) >
<\dot{X}^{\mu}_{L}(\zeta^{+}) ~\dot{\Pi}^{\nu}_{L}
(\tilde{\zeta}^{+}) > ={1\over 4} ~{D \over (\zeta^{+} -
\tilde{\zeta}^{+})^4}.
\eeqa
The contribution of the ghosts b-c system  to the
central charge remains the same as in the
standard bosonic string theory
$$
<T~T^{ghost}_{++}(\zeta^{+}) ~T^{ghost}_{++}(\tilde{\zeta}^{+})  >~ =~
-{13\over 4} ~{1 \over (\zeta^{+} -
\tilde{\zeta}^{+})^4}.
$$
The same is true for fermion and super-ghosts $\gamma - \beta$ system
\beqa
<T~T^{fermion}_{++}(\zeta^{+}) ~T^{fermion}_{++}(\tilde{\zeta}^{+})  >~ &+&~
<T~T^{super-ghost}_{++}(\zeta^{+}) ~T^{super-ghost}_{++}
(\tilde{\zeta}^{+})  >~ =~\nonumber\\= {1\over 16} ~{D \over (\zeta^{+} -
\tilde{\zeta}^{+})^4}~&+&~
{11\over 8} ~{1 \over (\zeta^{+} -
\tilde{\zeta}^{+})^4}.
\eeqa
therefore the absence of conformal anomaly
\be
{D\over 4} ~-~{13\over 4} ~+~{D\over 16} ~+~{11\over 8}~=0
\ee
requires that the
space-time should be 6-dimensional \footnote{We got the same
result using mode expansion, the details will be given elsewhere.}
\be
D_c = 6.
\ee
This result can be qualitatively understood if one take
into account the fact that the field equations here are of the forth order
and therefore we have two time more degrees of freedom in the bosonic
sector than in the standard string theory. It will be a subject of
another paper to investigate the spectrum of this theory in full details.

The author wish to thank J.Polchinski, E.Floratos, R.Manvelyan
and A.Kehagias for useful remarks and the Institute
for Theoretical Physics UCSB for kind hospitality
where part of this work was completed. This work
was supported in part by the by EEC Grant no. HPRN-CT-1999-00161.

\vfill

\begin{thebibliography}{99}
\bibitem{geo}G.K.Savvidy and K.G.Savvidy, Mod.Phys.Lett. A8
(1993) 2963\\G.K.Savvidy, JHEP 0009 (2000) 044\\
R.V.Ambartzumian and et al, Phys.Lett. B275 (1992) 99\\
G.K.Savvidy and K.G.Savvidy, Int.J.Mod.Phys. A8 (1993) 3993
\bibitem{polykov}A.M.Polyakov. Nucl.Phys.B268 (1986) 406\\
H.Kleinert. Phys.Lett. 174B (1986) 335\\
W.Helfrich. Z.Naturforsch. C28 (1973) 693; J.Phys.(Paris) 46 (1985) 1263\\
L.Peliti and S.Leibler. Phys.Rev.Lett. 54 (1985) 1690\\
D.Forster. Phys.Lett. 114A (1986) 115\\
T.L.Curtright and et.al. Phys.Rev.Lett. 57 (1986)799; Phys.Rev. D34 (1986) 3811\\
F.David. Europhys.Lett. 2 (1986) 577\\
P.O.Mazur and V.P.Nair. Nucl.Phys. B284 (1987) 146\\
E.Braaten and C.K.Zachos. Pys.Rev. D35 (1987) 1512\\
E.Braaten, R.D.Pisarski and S.M.Tye. Phys.Rev.Lett. 58 (1987) 93\\
P.Olesen and S.K.Yang. Nucl.Phys. B283 (1987) 73\\
R.D.Pisarski. Phys.Rev.Lett. 58 (1987) 1300
\bibitem{weingarten}D.Weingarten. Nucl.Phys.B210 (1982) 229 \\
T.Sterling and J.Greensite. Phys.Lett. B121 (1983) 345 \\
B.Durhuus,J.Fr\"ohlich and T.Jonsson. Nucl.Phys.B225 (1983) 183 \\
J.Ambj\o rn,B.Durhuus,J.Fr\"ohlich and T.Jonsson. Nucl.Phys.B290 (1987) 480\\
T.Hofs\"ass and H.Kleinert. Phys.Lett. A102 (1984) 420 \\
M.Karowski and H.J.Thun. Phys.Rev.Lett. 54 (1985) 2556\\
F.David. Europhys.Lett. 9 (1989) 575
\bibitem{manvelyan} R.~Manvelian and G.~Savvidy,
Phys.Lett.B533 (2002) 138
\bibitem{curtright}T.Curtright and P. van Nieuwenhuizen,
Nucl.Phys.B294 (1987)125
\bibitem{lastprep}G.K. Savvidy, Conformal invariant strings with extrinsic
curvature, hep-th 0202108
\bibitem{seiberg}R.~Gopakumar, S.~Minwalla, N.~Seiberg and A.~Strominger,
JHEP {\bf 0008}, 008 (2000)
\bibitem{ramond}P.Ramond, Phys.Rev.D3 (1971) 2415
\bibitem{neveu}A.Neveu and J.Schwarz, Nucl.Phys.B31 (1971) 86
\bibitem{zumino}J.L.Gervais and B.Sakita, Nucl.Phys.B34 (1971) 632\\
Y.Iwasaki and K.Kikkawa,Phys.Rev.D8 (1973) 440\\
B.Zumino, "Relativistic Strings and Supergauges"
pp.367-381 in "Renormalisation and Invariance in QFT" ed. E.Caianiello
(Plenum Press, 1974)\\
L.Brink,P. Di Vecchia and P.Howe, Phys.Lett.65B (1976) 471\\
S.Deser and B.Zumino, Phys.Lett.65B (1976) 369\\
A.M.Polyakov. Phys.Lett.103B (1981) 207; Phys.Lett.103B (1981) 211
\bibitem{green}M.B.Green, J.H.Schwarz and E.Witten, Superstring Theory.
Vol.1,2 Cambridge: Cambridge University Press (1997).
\bibitem{polchinski} J. Polchinski, String Theory. Vol.1,2
   Cambridge: Cambridge University Press (1998) .


\end{thebibliography}
\end{document}